\documentclass[prl,tighten,twocolumn]{revtex4}

\usepackage{graphicx}

\begin{document}

\hyphenation{nano-tube nano-tubes}

\title{Kinetics and mechanism of proton transport across membrane nanopores}

\author{Christoph Dellago}
\address{Faculty of Physics,
University of Vienna,                    
Boltzmanngasse 5, 1090 Vienna, Austria}

\author{Gerhard Hummer}
\address{Laboratory of Chemical Physics, Building 5, National Institute of
Diabetes and Digestive and Kidney Diseases, National Institutes of
Health, Bethesda, Maryland 20892-0520, U.S.A.}

\date{\today}

\vspace*{1cm}
\begin{abstract}
We use computer simulations to study the kinetics and mechanism of 
proton passage through a narrow--pore carbon--nanotube membrane 
separating reservoirs of liquid water. Free energy and 
rate constant calculations show that protons move across the membrane 
diffusively in single-file chains of hydrogen-bonded water molecules.
Proton passage through the membrane is opposed by a high barrier along
the effective potential, reflecting the large electrostatic penalty for
desolvation and reminiscent of charge exclusion in biological water channels.
At neutral pH, we estimate a translocation rate of about 1 proton per
hour and tube.
\end{abstract}

\maketitle

Long-range proton transfer is central to processes as diverse as
hydrogen fuel cells \cite{Paddison,Weber_Newman}, the enzymatic
function of many proteins, and in particular membrane biophysics
\cite{HilleBook}. To explore the fundamental question of water-mediated 
proton transfer, and to design robust proton conducting media for 
technological applications, studying simpler model systems is essential. The
quasi-one-dimensional water chains forming inside carbon nanotubes
\cite{GH_NATURE} have attracted considerable attention, with computer
simulations suggesting proton mobilities exceeding those even of bulk
water \cite{POMES_ROUX1,MEI,VOTH_WIRE,DellagoPRL,HallPRL,HassanJCP}. 
However, large conductivity requires in addition a high density of charge 
carriers, which depends on the free energy penalty required to remove 
protons from the bulk liquid and introduce them into the pores.
This then raises the question if water-filled nanotubes can actually 
carry protonic currents of high density, i.e., whether the electrostatic 
desolvation penalty of the proton is compensated, at least in part, by its 
exceptionally high mobility.

\begin{figure}[tb]
\includegraphics[width=5.5cm]{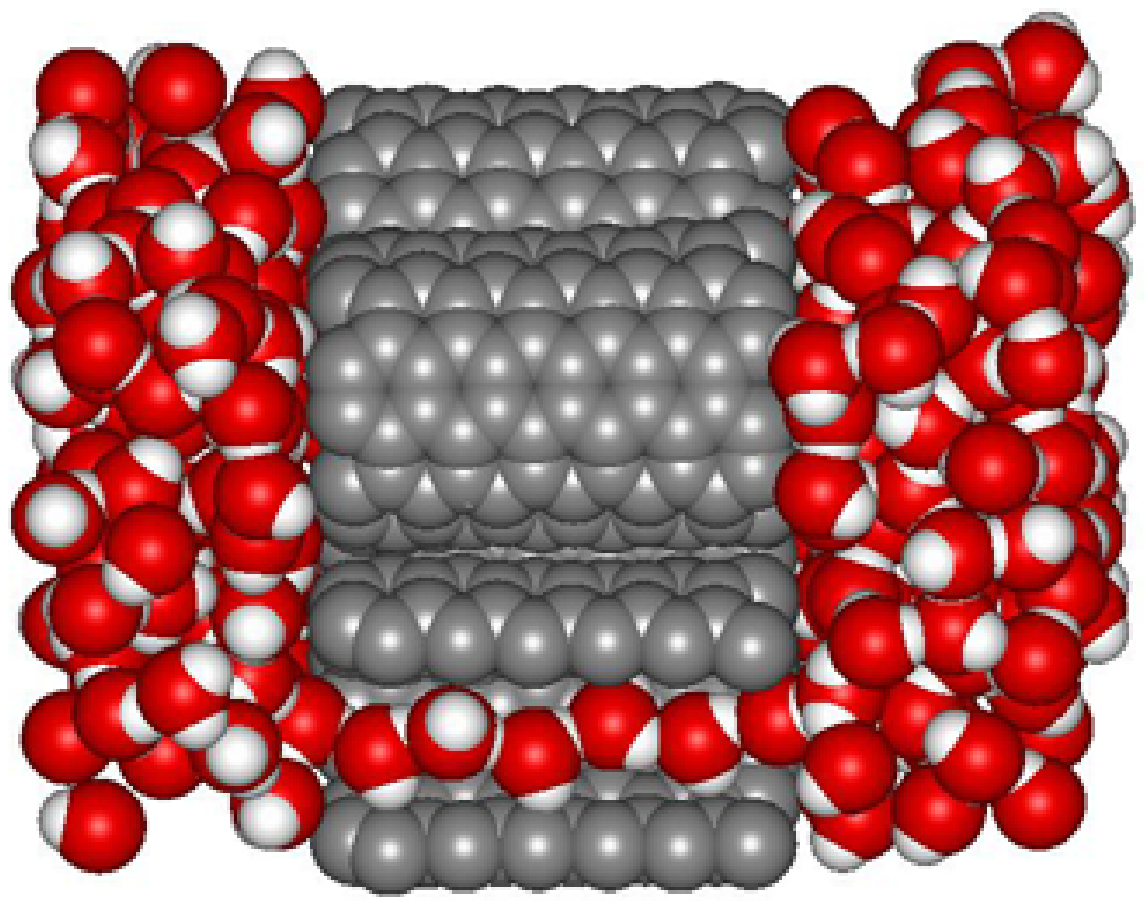}
\includegraphics[width=5.5cm]{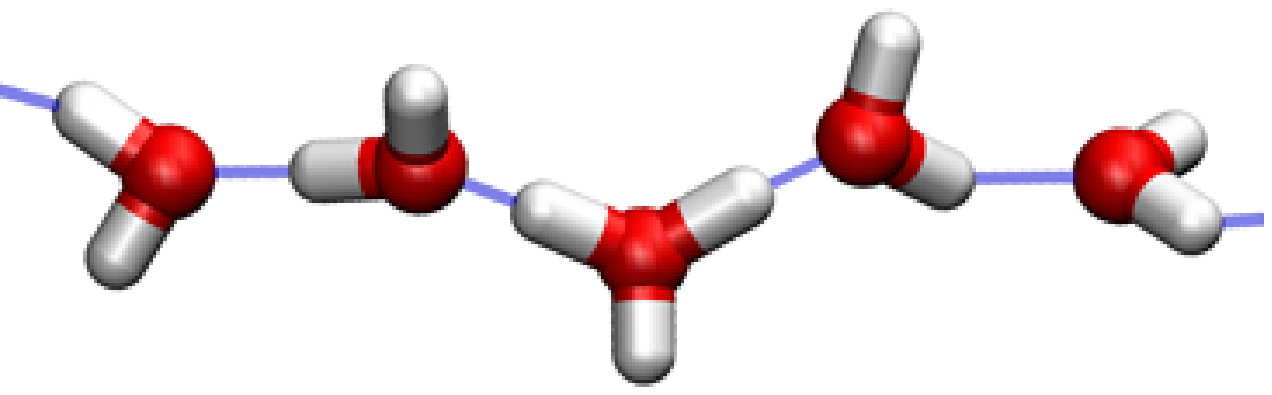}
\caption{(color online). {\em Top}: Side view of the carbon nanotube membrane 
immersed in liquid water. One carbon nanotube is cut open to expose the chain of 
hydrogen bonded water molecules traversing the pore. {\em Bottom:} Enlarged 
view of a typical configuration of a protonic defect in the water chain inside 
the pore.}
\label{fig:membrane}
\end{figure}

Here, we will use computer simulations to explore the kinetics and mechanism
of proton translocation through nanopores. In our simulations, four rigid
(6,6) armchair-type carbon nanotubes of 144 carbon atoms each are
packed into a hexagonal array to form a nanotube membrane in the 
periodically replicated simulation box (Fig. \ref{fig:membrane}).
The size of the simulation box in the $z$--direction parallel to the
tube axes is $34.3$ {\AA}, and $22.5$ {\AA} and $19.5$ {\AA},
respectively, in the $x$-- and $y$--directions. The  membrane is immersed 
in a bath of 292 water molecules containing one excess proton. At 
$T=300$ K and a density corresponding to that of liquid water, the 
$\sim$8-{\AA} diameter pores fill with single-file chains of six hydrogen 
bonded water molecules. In our simulations, the equations of motion 
are integrated with the velocity Verlet algorithm using a time step of 0.489 fs and a 
hydrogen mass of 2 a.m.u. For the interactions of the water
molecules and the excess proton, we use the multistate
empirical valence bond model (EVB) developed by Voth and collaborators 
\cite{VOTH} based on prior work of Warshel \cite{WARSHEL_EVB}.  This 
model accurately describes the energetics of bond breaking and 
formation during aqueous proton transfer and is computationally far 
less expensive than {\em ab initio} methodologies \cite{DellagoPRL}. The 
water oxygen atoms interact with the carbon atoms of the nanotube
through a Lennard-Jones potential with $\epsilon = 0.1143$ kcal/mol  
and $\sigma=3.27$ {\AA} yielding a channel aperture of approximately
2 {\AA}. Periodic boundary conditions with Ewald sums for the 
Coulombic interactions apply in all three spatial directions. 
We stress that the model and setup used here does not bias the simulation 
towards a particular H$^+$ transport mechanism.

In the bulk liquid outside the carbon nanotube membrane the excess 
proton moves primarily as a high mobility charge defect by proton transfers 
along the hydrogen bond network percolating through the liquid. 
During this so-called Grotthuss-process, the hydrated proton 
exists in a continuum of structures including as the limiting cases 
the Eigen cation H$_9$O$_4^+$, consisting of a hydronium ion 
H$_3$O$^+$ tightly hydrogen-bonded to three neighboring water 
molecules, and the Zundel cation  H$_5$O$_2^+$, in which the excess 
proton is shared between two water molecules
\cite{AGMON,MARX,AGMON_VOTH}. This structural diffusion process 
is rapid with typical proton hopping times on the picosecond timescale 
such that during the nanosecond simulations of this study each water 
molecule outside the membrane is visited several times by the excess 
charge. During these simulations, however, the proton never entered 
the membrane pores. 

\begin{figure}[tb]
\centerline{\includegraphics[width=7.0cm]{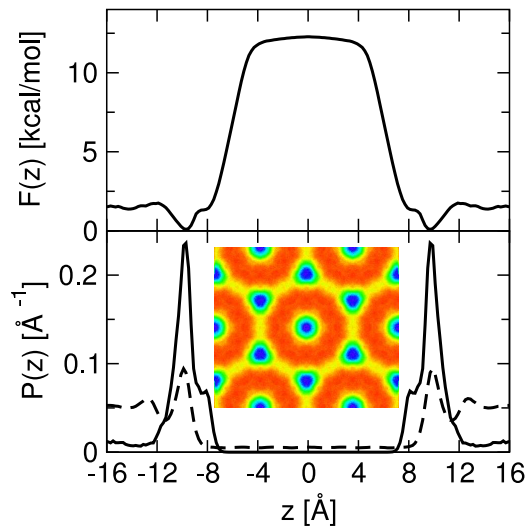}}
\caption{ \label{fig:Ftube} {\em Top:}
(color online). Free energy profile for the center of charge along $z$, 
where $z=0$ corresponds to the tube center.
{\em Bottom:} Corresponding probability distribution of the excess
charge (solid line) and a particular water oxygen (dashed line).
The inset shows the $xy$-projected probability density 
of the protonic defect in a slab $7{\rm \AA}<|z|<11{\rm \AA}$ just
outside the carbon nanotube membrane, obtained from a
molecular dynamics simulation of 2 ns. Red color indicates low proton
density and blue high proton density. The rims of the carbon nanotubes
are visible as red circles. }
\end{figure}

To clarify what prevents the proton from penetrating into the membrane 
interior despite the  high proton mobility along one-dimensional water
chains \cite{DellagoPRL} we have calculated the free energy profile $F(z)$ 
for the excess charge as a function of its position $z$ along the tube axis 
as shown in Fig. \ref{fig:Ftube}. We computed the free energy $F(z)$ 
inside the pore ($|z| \le 7.4$ {\AA}) using umbrella sampling 
Monte Carlo simulations in 10 separate windows. New configurations 
were generated with path sampling moves \cite{DBCC98,tps_review2}. 
In each window a total of 30,000 path shooting and shifting moves 
of 14.6 fs long trajectory segments were carried out amounting to a 
total simulation time of about 350 ps per window. Within the windows 
the free energy $F(z)$ was determined from the histogram $P(z)$ 
of the position of the center of charge \cite{EVB_NEW}, essentially 
the position of the hydronium ion averaged over all EVB states. The 
overall free energy profile was obtained by matching the free energies 
calculated in the separate windows and the 2.2~ns equilibrium run.

Coming from the bulk, the free energy $F(z)$ first decreases, goes through 
a minimum at $|z|\approx 10$ {\AA}, and then rises almost linearly,
reaching an approximately flat and 8 {\AA}-wide top near the tube center. 
The total free energetic cost of moving the excess charge from the bulk 
phase to the tube center is $\sim$10 kcal/mol, about 1/3 the cost for a 
sodium ion in a similar system \cite{Peter:BJ:2005}. As the motion of a proton along an 
isolated hydrogen bonded water chain is an essentially barrier-less 
process, this free energy penalty is due to the desolvation energy 
required to extract the proton from the favorable bulk environment 
and move it into the less polar interior of the pore, where the excess 
charge is coordinated by only two water molecules. Note, however, that 
the effective charge of the proton and hence also its desolvation 
penalty are substantially reduced by the dipolar polarization of the 
water chain \cite{DellagoPRL}, as discussed below. This effect is 
absent for other ionic species such as the sodium ion of Ref.~\onlinecite{Peter:BJ:2005}.

Whereas continuum electrostatics predicts the most favorable position
of the excess charge to be deep within the bulk liquid, the proton has 
an enhanced probability to be located near the apolar membrane (Fig. 
\ref{fig:Ftube} bottom). That the solvated proton is preferentially located 
near interfaces  has been observed earlier in simulations \cite{INTERFACE_SIM} 
and is consistent with experiments \cite{INTERFACE_EXP}. As depicted 
in the inset of Fig. \ref{fig:Ftube}, the excess charge appears to be 
located predominantly in two positions: either at the entrance of the 
C-nanotube or in the spaces between the nanotubes.  At both positions, 
the proton exists in its preferred Eigen-like configuration, in which the central 
hydronium ion donates three hydrogen bonds to water molecules, but 
accepts none. Such configurations occur also in the bulk liquid \cite{MARX}, 
but there they are less stable as they strain the hydrogen bond pattern of 
the surrounding liquid.

To study the mechanism and kinetics of the proton translocation process in 
detail we have carried out a rate constant calculation using the
reactive flux approach of Bennett and Chandler \cite{BENNETT,CHANDLER}. 
As a reaction coordinate we chose the position of the center of 
charge along the tube axis and we placed the dividing surface at the tube 
center perpendicular to its axis. A total of 5000 trajectories were initiated 
from initial conditions generated in a molecular dynamics simulation with a 
parabolic bias that kept the $z$-coordinate of the center of charge near the 
barrier top. The forces resulting from the bias on the center of charge were 
calculated with first order perturbation theory \cite{Dellago_Hummer_inprep}. 
An uncorrelated subset of the configuration with $z=0$ was then used as 
initial conditions for the reactive flux calculation. Initial momenta were drawn 
from an appropriate Maxwell-Boltzmann distribution.
Trajectories were terminated at $|z|=8$ {\AA}, from where the
probability of return to the dividing surface is negligible.

\begin{figure}[tb]
\centerline{\includegraphics[width=7.0cm]{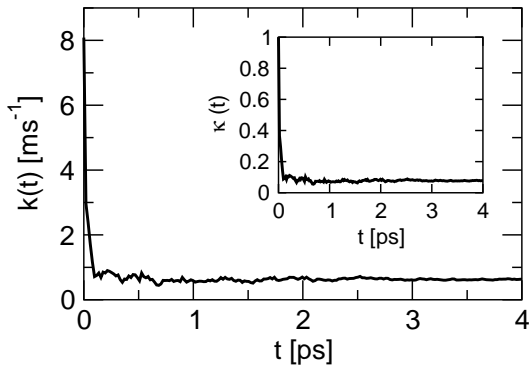} }
\caption{\label{fig:kappa}
Reactive flux $k(t)$ and transmission coefficient (inset).}
\end{figure}

The reactive flux  $k(t)$ calculated from
5000 trajectories is shown in Fig. \ref{fig:kappa}. The plateau value of 
$k(t)$  is the transmission rate constant $k\approx 6.4\times 10^2 {\rm s}^{-1}$ 
for one tube. For a proton concentration corresponding to pH=7 one 
obtains a protonic current of about 1 proton per hour and tube. We find that 
proton transport through the nanotube membrane is positively correlated with 
water flow. The corresponding electro-osmotic drag coefficient $K_{\rm drag}$ 
\cite{Paddison} is between about 0.5 and 1 water molecule per transported 
proton, as estimated from the correlated displacements of the proton and 
water chain in the reactive-flux simulations.
  
{From} the calculated rate $k$ of directional proton
translocations per pore in the absence of electric fields, one can
estimate the proton conductivity $\sigma$ of (6,6) nanotube membranes.
In the linear response limit, the number of protons translocated per
pore is $\approx k\beta eV$ where $V$ is the applied voltage and
$\beta=1/k_{B}T$. For an area density $\rho\approx 10^{18}$
m$^{-2}$ of nanotubes, the current density becomes $\rho k\beta
e^2\approx 100$ A m$^{-2} V^{-1}$ at room temperature for a rate
$k\approx 15$ s$^{-1}$ for pH$\approx$2.  This is about two orders of
magnitude below those of polymer electrolyte membranes used in fuel
cells \cite{Weber_Newman}.  However, this estimate ignores that the 
rate of proton translocation should here grow exponentially with applied 
voltage, as it is determined largely by proton desolvation (i.e., the low 
charge carrier concentration in the membrane) and not by the high 
proton mobility in the nanotubes.

As the protonic defect passes through the pore, it effectively
flips the dipolar orientation of the water chain.  This dipole
inversion is associated with a displacement current traveling in the
direction opposite to the proton motion.  As a consequence, the
effective charge transported through the membrane by proton
translocation alone is only about $\sim$60{\%} of an
elementary charge \cite{DellagoPRL}.
Proton transfer across the membrane is completed when the orientation of the
original dipole chain is restored by a hydrogen bonding defect passing through 
the pore \cite{POMES_ROUX1}.  This hydrogen bonding defect carries the remaining
$\sim$40{\%} of the elementary charge and its passage prepares the water chain 
for transport of the next proton. In separate simulations of a system of 4 
nanotubes and 292 TIP3P water molecules, we observed three 
reorientations during 15 ns, corresponding to a rate of about 1/(20 ns) per 
tube.  This is considerably slower than the rate of dipolar
reorientation in isolated tubes, $\sim$1/(2 ns)
\cite{GH_NATURE,Best_Hummer_PNAS_2005}, reflecting
the fact that reorientation proceeds through movement of a
hydrogen-bond defect that carries an effective charge through the
low-polarity membrane.  However, dipole reorientation is still much faster than
proton transfer, and thus not rate limiting.

The rather low transmission coefficient of $\kappa\approx 0.065$ (inset of 
Fig. \ref{fig:kappa}) found in our simulations may originate from two 
different causes. Either the  position of the center of charge is not a 
suitable reaction coordinate  capable of capturing the essential transition 
mechanism or the transition is of diffusive nature \cite{tps_review2}. In both 
cases frequent recrossings of the dividing surface reduce the transmission 
coefficient, albeit for very different reasons. We can distinguish these two 
cases by analyzing the trajectories started from the dividing surface. 
These trajectories were generated in pairs starting from the same configuration 
with momenta of identical magnitudes but opposite directions. In 54{\%} of all 
pairs the two trajectories reached different sides of the membrane and in 
46{\%} both trajectories relaxed to the same side. This result indicates that the 
forward and backward trajectories behave in an almost uncorrelated way as 
one would expect for diffusive barrier crossing \cite{Hummer_JCP_2004}. 
The average trajectory crosses the dividing surface at $z=0$
more than 8 times and there are trajectories which recross 60 times. 
This large number of recrossings is also indicative of diffusive 
dynamics.

\begin{figure}[tb]
\centerline{\includegraphics[width=7.0cm]{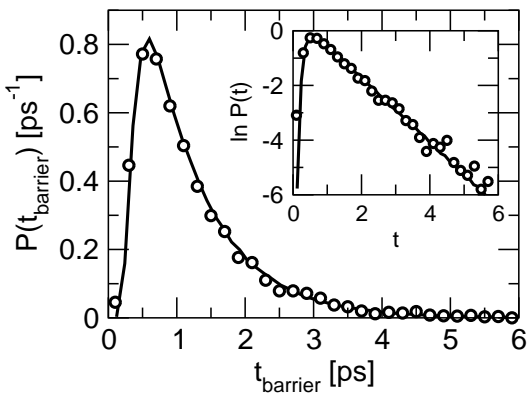}}
\caption{\label{fig:Tbarrier} Distribution of permanence times of the proton 
in the pore after release from the pore center from simulations
(circles) and Brownian diffusion on $F(z)$ (solid lines).
(Inset: logarithmic scale for $P(t)$.)
}
\end{figure}

To characterize the transition mechanism in more detail, we have analyzed
permanence times of the proton in the tube after release from $z=0$.        
The distribution of these permanence times extracted from 5000 
trajectories is shown in Fig. \ref{fig:Tbarrier}. Here, the permanence 
time on the barrier is the time the proton needs to reach $|z|=8$ {\AA}
starting from the barrier top. The distribution peaks at about 0.6ps and 
then decays exponentially  with a time constant of about 0.93ps. 
This distribution of permanence times is reproduced very well by 
a one-dimensional Brownian particle evolving on the effective
potential $F(z)$ of Fig. \ref{fig:Ftube} with a diffusion constant
$D=7$ {\AA}$^2$ps$^{-1}$, about half that 
estimated for protons in long water-filled tubes in
vacuum \cite{DellagoPRL}. The agreement between the molecular 
dynamics results for the full system and the one-dimensional Brownian 
dynamics simulation again indicates that the proton motion is diffusive 
and that the center of charge is an appropriate reaction coordinate.

The distribution of permanence times of the proton on the free energy barrier can 
be roughly modeled by a one-dimensional diffusion process on a flat
potential, starting from $z=0$ and terminated at $\pm L/2$.
The width of the almost flat barrier is $L\approx 8$ {\AA}
(Fig.~\ref{fig:Ftube}).  At long times, the resulting distribution of
permanence times decays as $P(t) \approx 4\pi D \exp(-\pi^2 Dt / L^2) / L^2$.
This exponential decay accurately reproduces the long time tail of the distribution 
of permanence times observed in our simulations and plotted in Fig. 
\ref{fig:Tbarrier}.

The resulting picture of a protonic defect diffusing through the pore
under the influence of the effective potential $F(z)$ has implications
on the design of conducting pores.  Increasing their length $L$ will 
reduce the transmission coefficient as $1/L$  \cite{RUIZ_MONTERO} and
hence lower the conductance. This effect will be enhanced by the larger
desolvation penalty  arising in longer pores. However, using nanopores of 
higher polarity, possibly embedded \cite{Bakajin:S:2006,Hinds:N:2005}
in a high-dielectric medium, should greatly reduce the desolvation 
cost and may result in the ideal combination of high proton mobility and 
concentration yielding proton current densities comparable to those measured for
polymer electrolyte membranes.


This work was supported by the Austrian Science Fund (FWF) under Grant
No. P17178-N02.  GH was supported by the Intramural Research Program
of the NIH, NIDDK.


\end{document}